\documentclass[preprint,preprintnumbers,amsmath,amssymb]{revtex4}
\usepackage{bm}% bold math
\begin{document}

\preprint{CAS-KITPC/ITP-087}

\title{~\\ \vspace{2cm}
K-essence Explains a Lorentz Violation Experiment\vspace{1cm}}

\author{Miao Li}\email{mli@itp.ac.cn}
\author{Yi Pang}\email{yipang@itp.ac.cn}
\author{Yi Wang}\email{wangyi@itp.ac.cn}
\affiliation{Kavli Institute for Theoretical Physics, Key Laboratory
of Frontiers in Theoretical Physics, Institute of Theoretical
Physics, Chinese Academy of Sciences, Beijing 100190, People's
Republic of China\vspace{2cm}
}%

\begin{abstract}
Recently, a state of the art experiment shows evidence for Lorentz
violation in the gravitational sector. To explain this experiment,
we investigate a spontaneous Lorentz violation scenario with a
generalized scalar field. We find that when the scalar field is
nonminimally coupled to gravity, the Lorentz violation induces a
deformation in the Newtonian potential along the direction of
Lorentz violation.
\end{abstract}

\pacs{11.30.Cp, 04.50.Kd, 04.80.Cc, 11.30.Qc}% PACS, the Physics and Astronomy
                             % Classification Scheme.

\maketitle

The pursuit of Lorentz violation has attracted increasing attention.
The local Lorentz symmetry has been examined in many sectors of the
standard model, including the sectors relating to photons,
electrons, protons, and neutrons \cite{AmelinoCamelia:2005qa,
Kostelecky:2003fs, Mattingly:2005re}. No Lorentz violation has been
identified so far in these sectors. The theoretical studies of
Lorentz violation can be found in \cite{Yan:1982tq,
Kostelecky:2003fs, Bailey:2006fd}.

Recently, M\"{u}ller, Chiow, Herrmann, Chu, and Chung
\cite{Muller:2007es} performed an experiment to probe the local
Lorentz symmetry in the gravitational sector. They measured the
phase shift of atoms using atom interferometry. Finally, they found
a more than 2$\sigma$ departure of Lorentz symmetry. This result may
be a signal of Lorentz symmetry violation.

In \cite{Muller:2007es}, the deviation from Lorentz symmetry is
parametrized in the SME (standard model extension) framework
\cite{Kostelecky:2003fs, Bailey:2006fd}. In SME, the Lorentz
violation originates from a Lorentz violating coupling in the action
\begin{equation}\label{a1}
S_{\rm{LV}}=\frac{M_p^2}{2} \int d^4 x \sqrt{-g}
(s^{\mu\nu}R^T_{\mu\nu}+t^{\mu\nu\alpha\beta}C_{\mu\nu\alpha\beta})~,
\end{equation}
where $s^{\mu\nu}$, $t^{\mu\nu\alpha\beta}$ indicate Lorentz
violation in gravity, $R^T_{\mu\nu}$ is the traceless part of
$R_{\mu\nu}$ and $C_{\mu\nu\alpha\beta}$ is the conformal Weyl
tensor. As a result of this coupling, $s^{\mu\nu}$ and
$t^{\mu\nu\alpha\beta}$ inherit the symmetries of the Ricci tensor
and the Riemann curvature tensor respectively.

The Lagrangian for a nonrelativistic test particle takes the form
\begin{equation}\label{pl}
  {\cal L}_p=\frac{1}{2}mv^2+G\frac{Mm}{r}\left( 1+
  \frac{s^{jk}r^j r^k}{2r^2}+\cdots \right)~,
\end{equation}
where ``$\cdots$'' denotes terms that are irrelevant to the
measurement.

If one assumes standard dispersion relation for photons, the Lorentz
violating tensor $s^{\mu\nu}$ is measured to be
\begin{equation}\label{lvexp}
  s^{XX}-s^{YY}=-(5.6 \pm 2.1) \times 10^{-9}~.
\end{equation}

In this article, we explain the anomaly in Eq. \eqref{lvexp} in
terms of scalar fields with a generalized kinetic term. We consider
the action
\begin{equation}\label{action}
  S=\int d^4x \sqrt{-g}\left( \frac{1}{2}M_p^2R+P(X)-
  \frac{1}{{\tilde M}^2}\partial^\mu\phi\partial^\nu\phi
  R_{\mu\nu}\right)~,
\end{equation}
where $X=-g^{\mu\nu}\partial_\mu\phi\partial_\nu\phi$, with the
signature of the metric $(-,+,+,+)$, and $\tilde M$ is an energy
scale denoting the coupling strength between $\phi$ and
$R_{\mu\nu}$. This kind of generalized kinetic term in the action
has been widely used in cosmology (see, {\it e.g.}
\cite{ArmendarizPicon:1999rj,Garriga:1999vw,ArkaniHamed:2003uy,Li:2008qc},
and references therein). For our purpose of breaking rotational
invariance, it is important that we have a negative vacuum
expectation value for $X$; thus $P(X)$ is a function of $X$ only.
This is guaranteed at the effective action level since we assume
that there is a shift symmetry of $\phi$: $\phi\rightarrow \phi+a$,
and this symmetry is respected also by the coupling between the
Ricci tensor and the gradient of $\phi$ as in Eq. (4). This action
has another feature that is to retain the shift symmetry of $\phi$,
$\partial^\mu\phi\partial^\nu\phi R_{\mu\nu}$ is the unique
nontrivial coupling between $\phi$, $R_{\mu\nu}$, and
$C_{\mu\nu\alpha\beta}$ when our discussion is just relevant up to
the first order derivative of $\phi$. The reason is that one cannot
construct $t^{\mu\nu\alpha\beta}$ from the gradient of $\phi$, since
$\partial^\mu\phi\partial^\nu\phi$ is symmetric about indices $\mu,
\nu$, while $t^{\mu\nu\alpha\beta}$ inheriting the symmetry of the
Riemann curvature tensor, is antisymmetric about indices $\mu, \nu$,
and $\alpha, \beta$. Thus we only need to introduce one parameter to
describe the interaction strength between $\phi$ and gravity in the
framework of SME. On the contrary, if the Lorentz violation is
induced by the vector field or tensor field \cite{ks1,
bkgrav,bkmass,bb,ms}; then one can check that up to the first order
derivative of these fields, their couplings to gravity will include
both the Ricci term and the Weyl term. Therefore to parametrize
these couplings, at least two parameters are necessary.

To have Lorentz violation in spacelike direction, the Hamiltonian
derived from the scalar field Lagrangian $P(X)$ must have a minima
at $X<0$. Note that the $X<0$ regime is an opposite limit compared
with the ghost condensation scenario, where $X$ has an expectation
value at $X>0$. As the expectation value of $X$ comes from
spontaneously breaking, our $X<0$ model has equal probability to be
realized compared with the ghost condensation scenario, so our model
has to be considered as seriously as ghost condensation. We also
find that our model has interesting cosmological implications, which
are to be discussed in a forthcoming work.

We use the Lagrangian
\begin{equation}
P(X)=\frac{X^2}{4M^4}+\frac{1}{2}X+\frac{1}{4}M^4~.
\end{equation}
To have the vacuum expectation value of $X$, we solve the equation
of motion of $\phi$ with $g_{\mu\nu}=\eta_{\mu\nu}$. The equation of
motion takes the form
\begin{equation}
  \partial_\mu\left(
  \sqrt{-g}\left(\frac{X}{M^4}+1\right)g^{\mu\nu}\partial_\nu
  \phi\right)=0~.
\end{equation}
Note that the solution $\partial_\nu \phi=0$ is not a stable
solution, since the corresponding energy is not minimal. The stable
solution of the above equation is $X=-M^4$, implying a Lorentz
violation.

Without losing generality, we assume the gradient of the scalar
field to be along the $z$ direction. We have $\partial_z \phi=M^2$.
Inspired by the experiment \cite{Muller:2007es}, the interaction
strength should take the form
\begin{equation}\label{alpha}
\alpha\equiv\frac{M^4}{M_p^2 \tilde M^2}\simeq 10^{-9}~.
\end{equation}

In the remainder of this article, we will show explicitly that our
model can explain the proposed Lorentz violation. To do this, we
will solve the perturbation equations with a point mass
$m\delta({\bf x})$. We find that the gravitational potential induced
by $m$ is indeed deformed in the $z$ direction. We also find that
the perturbation has positive mass squared, so the perturbation is
well defined and stable.

To consider perturbations, we let $\phi(x)=M^2 z+\pi(x)$ and the
perturbation of the metric
\begin{equation}
  ds^2=-(1+2\Phi(x))dt^2+(1-2\Psi(x))(dx^2+dy^2)+(1-2\tilde\Psi(x))dz^2~.
\end{equation}
The Einstein equation contains the following constraint equations
\begin{align}\label{constraint}
& \partial_0(M_p^2(\Psi+\tilde\Psi)+\frac{2M^2}{\tilde
    M^2}\chi)=0,\nonumber\\
 &  M_p^2(\tilde\Psi-\Phi)+\frac{2M^2}{\tilde M^2}\chi
  =0~,\nonumber\\
&  M_p^2(\Psi-\Phi)+\frac{2M^2}{\tilde M^2}\chi+\frac{2M^4}{\tilde
  M^2}\tilde \Psi-\frac{2M^4}{\tilde M^2}(\Psi-\Phi)=0~.
\end{align}
In the above equations, the first one can be rewritten as
\begin{equation}\label{solvecon}
      M_p^2(\Psi+\tilde\Psi)+\frac{2M^2}{\tilde M^2}\chi =\varphi~.
\end{equation}
where $\varphi\equiv \varphi({\bf
  x})$ is a function with no time dependence.

Inserting these constraint equations into the rest of Einstein
equation, we derive following two independent equations of motion:
\begin{align}\label{eom1}
&  \nabla^2\varphi-2\alpha\varphi_{,33}=m\delta({\bf
  x})~,
\\
& \frac{M^2}{\tilde M^2} \left(
-2\Psi_{,00}+\nabla^2(\Psi-\Phi)+\square\tilde\Psi+(\Psi-\tilde
\Psi)_{,33}\right) +\chi+M^2\tilde\Psi=0~,\label{eom2}
\end{align}
where $\chi\equiv \pi_{,3}$, $\square\equiv
-\partial_t^2+\partial_x^2$, and Eq. (\ref{eom2}) is consistent with
the equation of motion of $\pi$.

To solve above equations, we first express $\Phi$, $\Psi$, and
$\tilde \Psi$ in terms of $\varphi$ and $\chi$,
\begin{align}\label{sols}
  \Phi &=
  \frac{1-2\alpha}{1-3\alpha}\frac{\varphi}{2M_p^2}
  +\frac{1-2\alpha}{1-3\alpha}\frac{M^2}{\tilde M^2 M_p^2}\chi~,
  \nonumber\\
  \Psi &=\frac{1-4\alpha}{1-3\alpha}\frac{\varphi}{2M_p^2}
  -\frac{1-2\alpha}{1-3\alpha}\frac{M^2}{\tilde M^2 M_p^2}\chi~,
  \nonumber\\
  \tilde\Psi &=\frac{1-2\alpha}{1-3\alpha}\frac{\varphi}{2M_p^2}
  -\frac{1-4\alpha}{1-3\alpha}\frac{M^2}{\tilde M^2 M_p^2}\chi~.
\end{align}
In terms of $\varphi$ and $\chi$, Eq. \eqref{eom2} can be rewritten
as
\begin{equation}\label{eom0}
  -\square \chi-\frac{2\alpha}{3-8\alpha}\chi_{,33}+
  \frac{\tilde M^2}{2M^2(3-8\alpha)}\left(
\nabla^2\varphi-2\alpha\varphi_{,33}-4\alpha\nabla^2\varphi
  \right)+m_1^2\chi-m_2^2\varphi =0~,
\end{equation}
where
\begin{equation}
  m_1^2 \equiv \frac{(1-2\alpha)^2}{\alpha(3-8\alpha)} \tilde
  M^2~,\qquad m_2^2 \equiv \frac{1-2\alpha}{2(3-8\alpha)}\frac{\tilde M^4}{M^2}~.
\end{equation}

When we set $m=0$ in Eq. \eqref{eom1}, we have $\varphi=0$, and the
$\chi$ field has an oscillating solution with positive mass $m_1$,
much larger than $\tilde{M}$. In other words, the perturbation of
$\chi$ has positive mass squared. The perturbation is stable.

$\tilde{M}$ is the mass scale appearing in the Ricci term in Eq.
(4), representing the mass scale at which this term is generated. If
this term is due to quantum gravity effect, $\tilde{M}$ is close to
$M_p$. If for some purpose we want to have a much lower scale, we
will have to assume new physics (for instance a large extra
dimension) from which this Ricci term arises.

We further consider gravity with source. When $m\neq 0$, the
solution of Eq. \eqref{eom1} is
\begin{equation}\label{sol1}
  \varphi=-\frac{m}{\sqrt{1-2\alpha} 4\pi r'}~,\qquad r'^2\equiv
  x^2+y^2+z'^2~,\qquad z'\equiv \frac{z}{\sqrt{1-2\alpha}}~.
\end{equation}
Inserting Eq. \eqref{sol1} into Eq. \eqref{eom0}, we have
\begin{align}\label{chisol}
  \chi=&\frac{m\tilde M^2(1-4\alpha)}{2M^2(3-8\alpha)}\frac{e^{-m_1 r{''}}}{4\pi r{''}}
  -\frac{amm_2^2}{m_1\sqrt{1-2\alpha}}\int_0^1 \frac{1}{\sqrt{t}}
  e^{-\sqrt{t}m_1 r{'''}}dt
\nonumber\\&
  -\frac{4a\alpha^2}{(3-8\alpha)\sqrt{1-2\alpha}}
  \frac{m \tilde M^2}{m_1 M^2}\partial_3^2 \int_0^1\frac{1}{\sqrt{t}}
  e^{-\sqrt{t}m_1 r{'''}}dt~,
\end{align}
where
\begin{align}
  r{''}^2 &\equiv x^2+y^2+z{''}^2~,\qquad z{''}\equiv\frac{z}
  {\sqrt{1+2\alpha/(3-3\alpha)}}~,\nonumber\\
r{'''}^2 &\equiv x^2+y^2+z{'''}^2~,\qquad z{'''}\equiv\frac{z{''}}
  {\sqrt{t+(1-t)a^2}}~, \qquad a\equiv \sqrt{\frac{(1-2\alpha)(3-3\alpha)}
  {3-\alpha}}~.
\end{align}
To the first order in $\alpha$, Eq. \eqref{chisol} takes the form
\begin{equation}\label{chiapp}
  \frac{M^2}{\tilde M^2 M_p^2}\chi=\frac{1}{3}\frac{Gme^{-m_1r{''}}}{r{''}}
  -\frac{\alpha Gm}{r}\left( 1-e^{-m_1 r} \right)~.
\end{equation}
Note that the contribution from $\chi$ to the gravitation potential
$\Phi$ is either suppressed by the Yukawa factor $e^{-m_1 r{''}}$ or
by the small number $\alpha$. So at a long distance, the only
contribution from $\chi$ to $\Phi$ is a shift in the Newtonian
constant $G$. Insert Eqs. \eqref{sol1} and \eqref{chiapp} into Eq.
\eqref{sols}, and we have
\begin{equation}
  \Phi=-(1+3\alpha)\frac{Gm}{r}\left(1-\frac{z^2}{r^2}\alpha\right)~,
\end{equation}
where $G\equiv 1/(8\pi M_p^2)$, and the factor $1+3\alpha$ can be
absorbed into a redefinition of $G$, so it is not measurable.
Meanwhile the term $1-\frac{z^2}{r^2}\alpha$ gives an explicit
Lorentz violation. Comparing with \eqref{pl}, we have
\begin{equation}
  s^{33}=-2\alpha~.
\end{equation}

To compare with experiments, we can identify the third direction
(denoted by $z$ or 3 in the article) with the $X$ direction in
\cite{Muller:2007es}. Then a value $\alpha=2.8\times 10^{-9}$ gives
an explanation to the measurement \cite{Muller:2007es}.
Alternatively, we can also identify the $z$ direction with the $Y$
direction in \cite{Muller:2007es}, and let $\alpha=-2.8\times
10^{-9}$ to explain the experiment. In this case, $\tilde M^2<0$,
while the perturbation is still stable.

Theoretically, we find that physics at string scale may be
responsible for the small value of $\alpha$. The reason is below. As
mentioned before, it is reasonable to assume that the scalar-gravity
coupling term in Eq. (\ref{action}) originates from the quantum
gravity effects, so $\tilde{M}\simeq M_p$. Then from the expression
of $\alpha$ Eq. (\ref{alpha}), we read
\begin{equation}\label{}
    M=|\alpha|^{1/4}M_p\simeq1.7\times10^{16}\rm{GeV},
\end{equation}
where $M_p\simeq2.4\times10^{18}$GeV has been used ($M_p$ is the
reduced Planck mass defined as $M_p\equiv1/\sqrt{8\pi G}$). The
energy scale of $M$ can naturally arise from string theory by
requiring the scale of extra dimension approach Planck scale. At
this stage, we cannot guarantee the Lorentz violation is due to
stringy effects, but there is the possibility that Lorentz violation
is induced by some stringy physics effectively described by the
generalized scalar field.

Finally, we consider some signatures of our model at
small length scales. At small length scales, the first term in the
right hand side of Eq. (\ref{chiapp}) becomes important. Combining
this term and the contribution from $\varphi$, we will obtain a
gravitation potential with a running Newtonian constant,
\begin{equation}
    \Phi=-\frac{G(r)m}{r},
\end{equation}
and
\begin{equation}
    G(r)=(1-\frac{1}{3}e^{-m_1r})G,
\end{equation}
where we have neglected the terms proportional to $\alpha$. As we remarked before,
$\tilde{M}$ is the energy scale at which the Ricci term is generated, so naturally
it is not small, while $m_1$ is a factor $1/\alpha$ larger than $\tilde{M}$, an
even larger mass scale, so it is not conceivable to measure the running of the
Newton constant.

To conclude, we have considered spontaneous Lorentz violation from a
generalized scalar field. We show that when $X$ has a nonzero vacuum
expectation value, the Lorentz symmetry is spontaneously broken.
When coupling to gravity, this Lorentz violation affects the
Newtonian potential. This modification of gravity can explain the
current experiment \cite{Muller:2007es}, and can be tested in future
experiments. As stated in \cite{Muller:2007es}, future experiments
may reach the accuracy of $10^{-14}$. It is very interesting to see
whether the Lorentz violation is confirmed in the future.

\section*{Acknowledgments}
We thank M. L. Yan for discussion. This work was supported by NSFC
Grant No. 10525060, and a 973 project Grant No. 2007CB815401.


\begin{thebibliography}{99}

%\cite{AmelinoCamelia:2005qa}
\bibitem{AmelinoCamelia:2005qa}
  G.~Amelino-Camelia, C.~Lammerzahl, A.~Macias and H.~Muller,
  %``The search for quantum gravity signals,''
  AIP Conf.\ Proc.\  {\bf 758}, 30 (2005)
  [arXiv:gr-qc/0501053].
  %%CITATION = APCPC,758,30;%%


%\cite{Kostelecky:2003fs}
\bibitem{Kostelecky:2003fs}
  V.~A.~Kostelecky,
  %``Gravity, Lorentz violation, and the standard model,''
  Phys.\ Rev.\  D {\bf 69}, 105009 (2004)
  [arXiv:hep-th/0312310].
  %%CITATION = PHRVA,D69,105009;%%

%\cite{Mattingly:2005re}
\bibitem{Mattingly:2005re}
  D.~Mattingly,
  %``Modern tests of Lorentz invariance,''
  Living Rev.\ Rel.\  {\bf 8}, 5 (2005)
  [arXiv:gr-qc/0502097].
  %%CITATION = 00222,8,5;%%

%\cite{Yan:1982tq}
\bibitem{Yan:1982tq}
  M.~L.~Yan,
  %``The Renormalizability Of The General Gravity Theory With Torsion And The
  %Spontaneous Breaking Of Lorentz Group,''
  Commun.\ Theor.\ Phys.\  {\bf 2}, 1281 (1983).
  %%CITATION = CTPMD,2,1281;%%
%
%
%\cite{Bailey:2006fd}
\bibitem{Bailey:2006fd}
  Q.~G.~Bailey and V.~A.~Kostelecky,
  %``Signals for Lorentz violation in post-Newtonian gravity,''
  Phys.\ Rev.\  D {\bf 74}, 045001 (2006)
  [arXiv:gr-qc/0603030].
  %%CITATION = PHRVA,D74,045001;%%



%\cite{Muller:2007es}
\bibitem{Muller:2007es}
  H.~Muller, S.~W.~Chiow, S.~Herrmann, S.~Chu and K.~Y.~Chung,
  %``Atom Interferometry tests of the isotropy of post-Newtonian gravity,''
  Phys.\ Rev.\ Lett.\  {\bf 100}, 031101 (2008)
  [arXiv:0710.3768 [gr-qc]].
  %%CITATION = PRLTA,100,031101;%%


%\cite{ArmendarizPicon:1999rj}
\bibitem{ArmendarizPicon:1999rj}
  C.~Armendariz-Picon, T.~Damour and V.~F.~Mukhanov,
  %``k-Inflation,''
  Phys.\ Lett.\  B {\bf 458}, 209 (1999)
  [arXiv:hep-th/9904075].
  %%CITATION = PHLTA,B458,209;%%
%
%\cite{Garriga:1999vw}
 \bibitem{Garriga:1999vw}
  J.~Garriga and V.~F.~Mukhanov,
  %``Perturbations in k-inflation,''
  Phys.\ Lett.\  B {\bf 458}, 219 (1999)
  [arXiv:hep-th/9904176].
  %%CITATION = PHLTA,B458,219;%%
%
%\cite{ArkaniHamed:2003uy}
 \bibitem{ArkaniHamed:2003uy}
  N.~Arkani-Hamed, H.~C.~Cheng, M.~A.~Luty and S.~Mukohyama,
  %``Ghost condensation and a consistent infrared modification of gravity,''
  JHEP {\bf 0405}, 074 (2004)
  [arXiv:hep-th/0312099].
  %%CITATION = JHEPA,0405,074;%%
%
%\cite{Li:2008qc}
 \bibitem{Li:2008qc}
  M.~Li, T.~Wang and Y.~Wang,
  %``General Single Field Inflation with Large Positive Non-Gaussianity,''
  JCAP {\bf 0803}, 028 (2008)
  [arXiv:0801.0040 [astro-ph]].
  %%CITATION = JCAPA,0803,028;%%

\bibitem{ks1}
V.A.\ Kosteleck\'y and S.\ Samuel, Phys.\ Rev.\ D {\bf 40}, 1886
(1989).

\bibitem{bkgrav}
R.\ Bluhm and V.A.\ Kosteleck\'y, Phys.\ Rev.\ D {\bf 71}, 065008
(2005)

\bibitem{bkmass}
R.\ Bluhm, S.\ Fung, V.A.\ Kosteleck\'y, Phys.\ Rev.\ D {\bf 77},
065020 (2008)

\bibitem{bb}
 V.A.\ Kosteleck\'y and R.\ Lehnert, Phys.\ Rev.\ D {\bf 63},
065008 (2001); B.\ Altschul and V.A.\ Kosteleck\'y, Phys.\ Lett.\ B
{\bf 628}, 106 (2005); T.\ Jacobson and D.\ Mattingly, Phys.\ Rev.\
D {\bf 64}, 024028 (2001); C.\ Eling and T.\ Jacobson, Phys.\ Rev.\
D {\bf 69}, 064005 (2004); Class.\ Quant.\ Grav.\ {\bf 23}, 5643
(2006); P.\ Kraus and E.T.\ Tomboulis, Phys.\ Rev.\ D {\bf 66},
045015 (2002); S.M.\ Carroll and E.A.\ Lim, Phys.\ Rev.\ D {\bf 70},
123525 (2004); O.\ Bertolami and J.\ Paramos, Phys.\ Rev.\ D {\bf
72}, 044001 (2005); M.V.\ Libanov and V.A.\ Rubakov, JHEP {\bf
0508}, 001 (2005); J.W.\ Elliott {\it et al.}, JHEP {\bf 0508}, 066
(2005); R.\ Bluhm {\it et al.}, Phys.\ Rev.\ D {\bf 77}, 125007
(2008); S.M.\ Carroll {\it et al.}, Phys.\ Rev.\ D {\bf 79}, 065012
(2009), arXiv:0812.1050; {\bf 79}, 065011 (2009), arXiv:0812.1049.

\bibitem{ms}
M.D.\ Seifert, Phys.\ Rev.\ D {\bf 76}, 064002 (2007);
arXiv:0903.2279v1.

\end{thebibliography}
\end{document}